\pdfoutput=1
\documentclass[a4paper]{PoS}

\usepackage{amsmath}
\usepackage{mathtools}
\usepackage{color}

\allowdisplaybreaks

\newcommand\numberthis{\addtocounter{equation}{1}\tag{\theequation}}

\newcommand{\FORM}{{\sc Form}}
\newcommand{\SCHOONSCHIP}{{\sc Schoonschip}}
\newcommand{\Forcer}{{\sc Forcer}}
\newcommand{\Mincer}{{\sc Mincer}}

\newcommand{\prog}[1]{{\tt #1}}

\DeclareMathOperator{\NO}{{\bf NO}}
\DeclareMathOperator{\IBP}{{\bf IBP}}
\newcommand{\ep}{\epsilon}
\newcommand{\GOOD}[1]{{\color{blue} #1}}
\newcommand{\BAD}[1]{{\color{red} #1}}

\title{Forcer: a FORM program for 4-loop massless propagators}

\ShortTitle{Forcer: a FORM program for 4-loop massless propagators}

\author{
  \speaker{T.~Ueda},$^a$
  B.~Ruijl$^{ab}$
  and J.A.M. Vermaseren$^a$ \\
  \llap{$^a$}
  Nikhef Theory Group,
  Science Park 105, 1098 XG Amsterdam, The Netherlands \\
  \llap{$^b$}
  Leiden Centre of Data Science,
  Leiden University,
  Niels Bohrweg 1, 2333 CA Leiden, The Netherlands \\
  E-mail:
  \email{tueda@nikhef.nl},
  \email{benrl@nikhef.nl},
  \email{t68@nikhef.nl}
}

\abstract{
  We present a new \FORM{} program for analytically evaluating four-loop
  massless propagator-type Feynman integrals in an efficient way.
  Our program \Forcer{} implements parametric reductions of the aforementioned
  class of Feynman integrals into a set of master integrals and can be
  considered as a four-loop extension of the three-loop \Mincer{} program.
  Since the program structure at the four-loop level is highly complicated
  and the equations easily become lengthy, most of the code was generated in an
  automatic way or with computer-assisted derivations.
  We have checked correctness of the program by recomputing already-known
  quantities in the literature.
}

\FullConference{
  Loops and Legs in Quantum Field Theory \\
  24-29 April 2016 \\
  Leipzig, Germany
  \hfill
  \begin{picture}(0,0)
    \put(-80,585){\normalfont NIKHEF-2016-036}
  \end{picture}
}

\begin{document}


\section{Introduction}

Evaluating multi-loop massless self-energy diagrams and related quantities is
one of the typical problems in computing higher-order corrections to physical
observables (see ref.~\cite{Baikov:2015tea} for a recent review).
The \Mincer{} program, firstly implemented~\cite{Gorishnii:1989gt} in
\SCHOONSCHIP{}~\cite{Strubbe:1974vj} and later reprogrammed~\cite{Larin:1991fz}
in \FORM{}~\cite{Kuipers:2012rf}, was made for calculations of such integrals
up to the three-loop level and has been used in many phenomenological
applications.

The algorithm in the \Mincer{} program is based on the following facts:
\begin{itemize}
  \item A large part of integrals can be reduced into simpler ones via
    integration-by-parts identities (IBPs)~%
        \cite{Tkachov:1981wb,Chetyrkin:1981qh}
        of dimensionally regularized~\cite{Bollini:1972ui,'tHooft:1972fi}
        Feynman integrals.
        In particular, the so-called triangle rule, for which an explicit
        summation formula was given in ref.~\cite{Tkachov:1984xk},
        can be applied to many of the cases in the class.
  \item One-loop massless integrals are easily performed and expressed in
        terms of G-functions~\cite{Chetyrkin:1980pr,Chetyrkin:1981qh}.
        This also gives reductions of multi-loop integrals into those with
        one loop less but with non-integer powers of propagators.
\end{itemize}
After the whole reduction procedure, all integrals are expressed in terms of
the gamma functions and two master integrals. Laurent series expansions of the
latters with respect to the regulator $\epsilon$ were obtained by
another method~\cite{Chetyrkin:1980pr}, where $D=4-2\epsilon$ is the number of
space-time dimensions.

To perform massless propagator-type Feynman integrals at the four-loop level,
there exist generic and systematic ways of the IBP reductions; these include:
Laporta's algorithm~\cite{Laporta:2001dd} (see~%
\cite{%
Anastasiou:2004vj,%
Smirnov:2008iw,%
Smirnov:2013dia,%
Smirnov:2014hma,%
Studerus:2009ye,%
vonManteuffel:2012np%
} for public implementations),
Baikov's method~\cite{Baikov:1996rk,Baikov:2005nv} and heuristic search of the
reduction rules~\cite{Lee:2012cn,Lee:2013mka}.
After the reduction to a set of master integrals, one can substitute
the Laurent series expansions of the master integrals given in
refs.~\cite{Baikov:2010hf,Lee:2011jt}.
Those IBP solvers work for the four-loop calculations;
however, the IBP reduction usually takes much time and can easily become
the bottleneck of the computation.
More efficient reduction programs are demanded especially when one considers
extremely time-consuming calculations, for example, higher moments of the
four-loop splitting functions and the Wilson coefficients in deep-inelastic
scattering~\cite{Ruijl:2016pkm}.

This work aims to develop a new \FORM{} program \Forcer{}~%
\cite{Ueda:2016sxw,forcer}, which is specialized for the four-loop massless
propagator-type Feynman integrals and must be more efficient than using general
IBP solvers.
This is achieved by extending the algorithm of the three-loop \Mincer{} program
into the four-loop level.
As the program can be highly complicated and error-prone at the four-loop level,
the program should be generated in as automatic a way as possible, rather than
coded by hand.

We specifically emphasize the following points that require automatization:
\begin{itemize}
  \item Classification of topologies and constructing the reduction flow.
        All topologies appear at the four-loop level are enumerated and their
        graph structures are examined.
        The information is used for constructing the reduction flow from
        complicated topologies to simpler ones.
  \item Derivation of special rules.
        When substructures in a topology do not immediately give a
        reduction into simpler ones, one has to solve IBPs to obtain reduction
        rules for the topology, which removes at least one of the propagators
        or leads to a reduction into master integrals in the topology.
        They are obtained as symbolic rules, i.e., we allow the rules to contain
        indices (powers of the propagators and irreducible numerators) of
        the integrals as parameters.
\end{itemize}
The former can be fully automated by representing topologies as graphs in
graph theory and utilizing several graph algorithms.
In contrast, the latter is not trivial particularly when one considers
the automatization.
Much effort has been made for systematic search of symbolic rules using
the Gr\"obner basis~\cite{Tarasov:1998nx,Gerdt:2004kt} and the so-called
s-basis~\cite{Smirnov:2006wh,Smirnov:2006tz} as well as heuristics methods~%
\cite{Lee:2012cn,Lee:2013mka}.
Our approach for the special rules is more or less similar to ones in refs.~%
\cite{Lee:2012cn,Lee:2013mka}.
However, we constructed the reduction schemes for the special topologies
with human intervention on top of computer-assisted derivations,
often by trial and error for obtaining more efficient schemes.
This was possible because in the \Mincer{}/\Forcer{} approach many topologies
are reducible by rules derived from their substructures and we need special
rules only for a limited number of topologies.


\section{Constructing the reduction flow for all topologies}

\begin{figure}[t]
  \centering
  \begin{tabular}{cccc}
    \raisebox{5pt}{\includegraphics{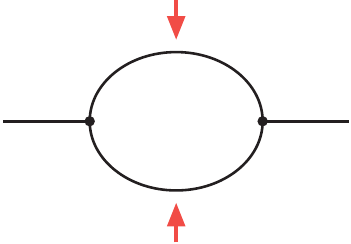}} &
    \raisebox{5pt}{\includegraphics{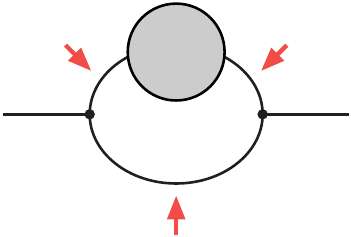}} &
    \raisebox{5pt}{\includegraphics{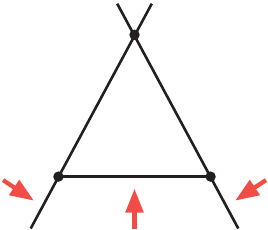}} &
    \raisebox{0pt}{\includegraphics{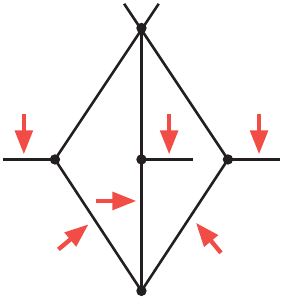}} \\
    (a) & (b) & (c) & (d)
  \end{tabular}
  \caption{
    Substructures of topologies leading to reduction into simpler ones:
    (a) one-loop insertion,
    (b) one-loop carpet,
    (c) triangle and
    (d) diamond.
  }
  \label{fig:substruct}
\end{figure}

In order to manage many topologies appearing at the four-loop level, we need to
automatize classification of topologies and constructing the reduction flow.
We represent each topology as an undirected graph in graph theory,
which makes it easy to detect the following types of substructures
in a topology, depicted in figure~\ref{fig:substruct}, by pattern matchings of
connections among vertices and edges:
\begin{enumerate}
  \item[(a)] One-loop insertion: the lines indicated by the arrows are
             integrated out.
  \item[(b)] One-loop carpet: the lines indicated by the arrows are
             integrated out.
  \item[(c)] Triangle: the triangle rule removes one of the three lines
             indicated by the arrows.
  \item[(d)] Diamond: the diamond rule~\cite{Ruijl:2015aca} removes one of the
             six lines indicated by the arrows.
\end{enumerate}
When none of the above is available, one has to solve the IBPs so as to obtain
special rules.

Starting from the top-level topologies that consist of 3-point vertices,
we consider removing a line from each topology in all possible ways.
Massless tadpole topologies are immediately discarded.
When a one-loop insertion/carpet is available in a topology,
the loop integration can remove more than one line at a time,
which should also be taken into consideration.
For example, removing a line from the two lines in the one-loop topology gives
a massless tadpole to be discarded, while removing both the two lines gives
the Born graph.
Some of generated topologies are often identical under graph isomorphism,
which is efficiently detected by graph algorithms.

Irreducible numerators of a topology have to be chosen such that they do not
interfere with the reduction determined by the topology substructure.
Conversely, two topologies may have in general totally different sets of
irreducible numerators, even if one is a derived topology obtained from
the other by removing some of the lines and they have many common propagators.
This means that a transition from a topology to another requires rewriting of
irreducible numerators, which can be a bottleneck of computing complicated
integrals with high powers of numerators.

Repeating this procedure until all topologies are reduced into the Born
graph gives a reduction flow of all possible topologies.
For the implementation, we used \prog{Python} with a graph library
\prog{igraph}~\cite{igraph}.
From 11 top-level topologies at the four-loop level, we obtained 437 non-trivial
topologies in total.
The numbers of topologies containing one-loop insertions, one-loop carpets,
triangles and diamonds are 335, 24, 53 and 4, respectively.
The remaining 21 topologies require construction of special rules.


\section{Derivation of special rules}

We start from the set of IBPs constructed in the usual way:
\begin{equation}
  S_0 = \{ I_1, \cdots, I_{R_0} \} ,
\end{equation}
where each IBP is given as $I_i = 0$ and the number of the relations is
$R_0=L(L+E)$ for an $L$-loop topology with $E$-independent external momenta.
We also define a set $S_1$ of IBPs obtained from $S_0$ with an index shifted
by one in all possible ways.
The set $S_1$ has $R_1 = N R_0$ elements for a topology with $N$-indices.
The set $S_2$ can be defined from $S_0$ by shifting an index by two or two
indices by one. The sets $S_3,S_4,\dots$ may be defined in a similar way,
but the number of the elements grows rather rapidly.

Then we consider searching useful reduction rules given by a linear combination
of IBPs in the combined set of $S_0\cup S_1$:
\begin{equation}
  \sum_i c_i I_i = 0, \qquad I_i \in S_0 \cup S_1 ,
  \label{eq:rule}
\end{equation}
by trying to eliminate complicated integrals in the system.
This may sound like the Laporta's algorithm: defining sets of IBPs
$S_0,S_1,\dots$ corresponds to generation of equations by choosing so-called
seeds in the Laporta's algorithm.
Note that, however, we keep indices of integrals as parameters,
unless we have already applied recursions to bring some of the indices
down to fixed values (usually one or zero).
The coefficients $c_i$ in eq.~\eqref{eq:rule} may contain the indices as
parameters.
Although one could expect that including sets with higher shifts
$S_2,S_3,S_4,\dots$ makes the system overdetermined at some point,
it never works because the coefficients can become immoderately complicated
in an intermediate step of solving the system.
Nevertheless, we observed that, with careful selection of linear combinations
and of the order of reducing indices, in many cases the combined set
$S_0\cup S_1$ is enough for obtaining reasonable rules and the complexity of
coefficients in the rules can be under control.

A combination of reduction rules gives a reduction scheme of a topology into
master integrals or simpler topologies.
One needs criteria to select rules as components of a scheme, for example,
\begin{itemize}
  \item Which index is reduced to a fixed value: bringing an index that
        entangles other indices in a complicated way
        down to a fix value often makes the reduction easy afterwards.
  \item Reducing complexity of integrals: it is ideal that recursion rules
        reduce the complexity of every integral at each step, which
        automatically guarantees the termination of the recursions.
  \item The number of terms generated by rules: a large number of generated
        terms in recursions are very stressful for computer algebra systems.
  \item Complexity of coefficients: as one uses exact rational arithmetic,
        complicated denominators of coefficients in generated terms considerably
        slow down the calculation.
        Moreover, spurious poles, which appear in denominators at intermediate
        steps but cancel out in the final result, can be problematic
        when one wants to expand and truncate the coefficients
        instead of using exact rational arithmetic during the calculation.
\end{itemize}
In practice, it is often difficult to find rules that satisfy all of the above
criteria and one has to be reconciled to unsatisfactory rules at some extent.
It is also true that predicting efficiency of a rule from its expression
is non-trivial.
Efficiency of reduction rules should be judged by how much it affects
the total performance in adequate benchmark tests
for the whole reduction scheme.

\begin{figure}[t]
  \centering
  \includegraphics{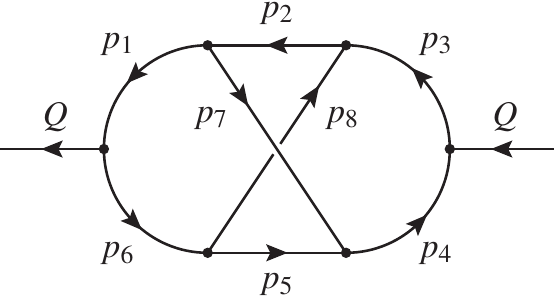}
  \caption{The three-loop non-planar topology.}
  \label{fig:no}
\end{figure}


\subsection{Example: special rules for the three-loop non-planar topology}

As an example, let us consider the three-loop non-planar topology
(figure~\ref{fig:no}):
\begin{equation}
  \NO(n_1,\dots,n_9) = \int d^D p_1 d^D p_2 d^D p_3
  \frac{(Q\cdot p_2)^{-n_9}}{(p_1^2)^{n_1} \dots (p_8^2)^{n_8}} .
\end{equation}
The indices for the propagators take positive integer values:
$n_1 \ge 1, \dots, n_8 \ge 1$, while the index $n_9$ for the irreducible
numerator is negative: $n_9 \le 0$.
When at least one of the indices for the propagators turns into a non-positive
integer, the integral becomes one with a simpler topology.
For simplicity, we set $Q^2=1$.
As the set $S_0$, we have 12 IBPs, which are schematically given by
\begin{align}
\begin{aligned}
  I_1 &= \IBP\left(\frac{\partial}{\partial p_1}\cdot p_1,\NO\right) , &
  I_2 &= \IBP\left(\frac{\partial}{\partial p_1}\cdot p_2,\NO\right) , &
  I_3 &= \IBP\left(\frac{\partial}{\partial p_1}\cdot p_3,\NO\right) ,
  \\
  I_4 &= \IBP\left(\frac{\partial}{\partial p_1}\cdot Q,\NO\right) , &
  I_5 &= \IBP\left(\frac{\partial}{\partial p_2}\cdot p_1,\NO\right) , &
  I_6 &= \IBP\left(\frac{\partial}{\partial p_2}\cdot p_2,\NO\right) ,
  \\
  I_7 &= \IBP\left(\frac{\partial}{\partial p_2}\cdot p_3,\NO\right) , &
  I_8 &= \IBP\left(\frac{\partial}{\partial p_2}\cdot Q,\NO\right) , &
  I_9 &= \IBP\left(\frac{\partial}{\partial p_3}\cdot p_1,\NO\right) ,
  \\
  I_{10} &= \IBP\left(\frac{\partial}{\partial p_3}\cdot p_2,\NO\right) , &
  I_{11} &= \IBP\left(\frac{\partial}{\partial p_3}\cdot p_3,\NO\right) , &
  I_{12} &= \IBP\left(\frac{\partial}{\partial p_3}\cdot Q,\NO\right) .
\end{aligned}
\end{align}
We also build the set $S_1$ of 108 IBPs from $S_0$ by shifting $n_i\to n_i-1$
for all possible ways.

We solve the system consisting of $S_0 \cup S_1$ by eliminating complicated
integrals.
First, we take a linear combination of IBPs as
\begin{equation}
\begin{split}
  \MoveEqLeft
       2 I_1
       + 4 I_6
       + 2 I_{11}
       - I_2(n_9\to n_9+1)
       - 2 I_6(n_9\to n_9+1)
       - I_{10}(n_9\to n_9+1)
  \\&
       + I_2(n_1\to n_1-1,n_9\to n_9+1)
       - I_4(n_1\to n_1-1,n_9\to n_9+1)
  \\&
       - I_4(n_2\to n_2-1,n_9\to n_9+1)
       - 2 I_8(n_2\to n_2-1,n_9\to n_9+1)
  \\&
       - I_{12}(n_2\to n_2-1,n_9\to n_9+1)
       + I_{10}(n_3\to n_3-1,n_9\to n_9+1)
  \\&
       - I_{12}(n_3\to n_3-1,n_9\to n_9+1)
       - I_{10}(n_4\to n_4-1,n_9\to n_9+1)
  \\&
       - I_2(n_6\to n_6-1,n_9\to n_9+1)
       + I_4(n_7\to n_7-1,n_9\to n_9+1)
  \\&
       + I_{12}(n_8\to n_8-1,n_9\to n_9+1)
  = 0 ,
  \label{eq:35}
\end{split}
\end{equation}
where $I_i(n_j \to n_j \pm 1,\dots)$ represents $I_i$ with the indices shifted.
We have also globally shifted $n_9\to n_9+1$ in $S_0\cup S_1$.
The linear combination in eq.~\eqref{eq:35} gives a rule
\begin{equation}
\begin{split}
  \MoveEqLeft
  \NO(n_1,n_2,n_3,n_4,n_5,n_6,n_7,n_8,n_9)
  =
  \frac{1}{2(n_1+n_2+n_3+n_4+n_5+n_6+n_7+n_8+n_9-7+4\ep)}
  \Bigl[
    \\& + n_1 \NO(\BAD{n_1+1},\GOOD{n_2-1},n_3,n_4,n_5,n_6,n_7,n_8,\GOOD{n_9+1})
    \\& - n_1 \NO(\BAD{n_1+1},n_2,n_3,n_4,n_5,n_6,\GOOD{n_7-1},n_8,\GOOD{n_9+1})
    \\& + n_3 \NO(n_1,\GOOD{n_2-1},\BAD{n_3+1},n_4,n_5,n_6,n_7,n_8,\GOOD{n_9+1})
    \\& - n_3 \NO(n_1,n_2,\BAD{n_3+1},n_4,n_5,n_6,n_7,\GOOD{n_8-1},\GOOD{n_9+1})
    \\& + (n_1+2n_2+n_3+n_9-3+2\ep) \NO(n_1,n_2,n_3,n_4,n_5,n_6,n_7,n_8,\GOOD{n_9+1})
    \\& + (n_9+1) \NO(n_1,\GOOD{n_2-1},n_3,n_4,n_5,n_6,n_7,n_8,\GOOD{n_9+2})
  \Bigr] ,
\end{split}
\label{eq:36}
\end{equation}
which increases $n_9$ at least by one and can be repeatedly applied until
$n_9$ becomes $0$.
It also decreases the total complexity of integrals $\sum_{i=1}^8 n_i - n_9$.
The sum $\sum_{i=1}^9 n_i$ appearing in the denominator is monotonically
increased by one,
therefore the coefficients get a single pole at most in the recursion.
The last term increases $n_9$ by two, but due to its coefficient $(n_9+1)$,
which vanishes when $n_9 = -1$, it is guaranteed that $n_9$ never turns into
a positive value.

Next, we set $n_9 = 0$ in the system and take a linear combination
\begin{equation}
  \Bigl[
  - I_1(n_1\to n_1-1)
  + I_4(n_1\to n_1-1)
  - I_6(n_1\to n_1-1)
  + I_7(n_1\to n_1-1)
  \Bigr]
  \Bigr|_{n_9=0}
  = 0 ,
\end{equation}
which leads to
\begin{equation}
\begin{split}
  \MoveEqLeft
  \NO(n_1,n_2,n_3,n_4,n_5,n_6,n_7,n_8,0)
  =
  \frac{1}{n_1-1}
  \Bigl[
    \\& - n_7 \NO(\GOOD{n_1-1},n_2,n_3,\GOOD{n_4-1},n_5,n_6,\BAD{n_7+1},n_8,0)
    \\& + n_7 \NO(\GOOD{n_1-1},n_2,n_3,n_4,\GOOD{n_5-1},n_6,\BAD{n_7+1},n_8,0)
    \\& - n_2 \NO(\GOOD{n_1-1},\BAD{n_2+1},\GOOD{n_3-1},n_4,n_5,n_6,n_7,n_8,0)
    \\& + n_2 \NO(\GOOD{n_1-1},\BAD{n_2+1},n_3,n_4,n_5,n_6,n_7,\GOOD{n_8-1},0)
    \\& + (n_1+n_2+2n_5+2n_6+n_7+2n_8-9+4\ep) \NO(\GOOD{n_1-1},n_2,n_3,n_4,n_5,n_6,n_7,n_8,0)
    \\& + (n_1-1) \NO(n_1,n_2,n_3,n_4,n_5,\GOOD{n_6-1},n_7,n_8,0)
  \Bigr] .
\end{split}
\label{eq:38}
\end{equation}
This rule decreases the total complexity as well as $(n_1+n_6)$
and is applicable unless $n_1 = 1$.
In fact, it is a variant of the diamond rule~\cite{Ruijl:2015aca} and
its existence is expected from the substructure of the topology.
Similar rules exist for $n_3$, $n_4$ and $n_6$ due to the flipping symmetry of
the topology. Eqs.~\eqref{eq:36} and~\eqref{eq:38} were used also in
the original \Mincer{}.

Lastly, we set $n_1=1$, $n_3=1$, $n_4=1$, $n_6=1$ and $n_9=0$.
Then a solution of the system is obtained by the following linear combination:
\begin{equation}
\begin{split}
  \MoveEqLeft
  \Bigl[
         (n_2 - 1) I_2
       + (n_2 - 1) I_6
       + (n_2 - 1) I_9
       - (n_2 - 1) I_{10}
       + (n_2 - 1) I_{11}
       - (n_2 - 1) I_{12}
  \\&
       + 2 (n_2 - 1) I_1(n_1\to n_1-1)
       - (n_2 - 1) I_2(n_1\to n_1-1)
       - (n_2 - 1) I_4(n_1\to n_1-1)
  \\&
       + (n_2 - 1) I_6(n_1\to n_1-1)
       - (n_2 - 1) I_8(n_1\to n_1-1)
       + (n_2 - 1) I_{11}(n_1\to n_1-1)
  \\&
       - (n_2 - 1) I_{12}(n_1\to n_1-1)
       + 2 (n_2 - 1) I_1(n_2\to n_2-1)
       - (n_2 - 1) I_4(n_2\to n_2-1)
  \\&
       - 2 (n_2 + 2 n_5 + 2 n_7 + n_8 - 6 + 4 \ep) I_5(n_2\to n_2-1)
  \\&
       + 2 (2 n_2 + 2 n_5 + 2 n_7 + n_8 - 7 + 4 \ep) I_6(n_2\to n_2-1)
  \\&
       - 2 (2 n_2 + 2 n_5 + 2 n_7 + n_8 - 7 + 4 \ep) I_7(n_2\to n_2-1)
  \\&
       + (3 n_2 + 4 n_5 + 4 n_7 + 2 n_8 - 13 + 8 \ep) I_8(n_2\to n_2-1)
  \\&
       - (n_2 - 1) I_9(n_3\to n_3-1)
       + (n_2 - 1) I_{10}(n_3\to n_3-1)
       - 2 (n_2 - 1) I_{11}(n_3\to n_3-1)
  \\&
       + 2 (n_2 - 1) I_{12}(n_3\to n_3-1)
       - (n_2 - 1) I_9(n_4\to n_4-1)
       + (n_2 - 1) I_{10}(n_4\to n_4-1)
  \\&
       + (n_2 - 1) I_{12}(n_4\to n_4-1)
       + 2 (n_2 - 1) I_{11}(n_5\to n_5-1)
       - (n_2 - 1) I_{12}(n_5\to n_5-1)
  \\&
       - 2 (n_2 - 1) I_1(n_6\to n_6-1)
       - (n_2 - 1) I_2(n_6\to n_6-1)
       - 3 (n_2 - 1) I_6(n_6\to n_6-1)
  \\&
       - (n_2 - 1) I_{11}(n_6\to n_6-1)
       + (n_2 - 1) I_{12}(n_6\to n_6-1)
       - 2 (n_2 - 1) I_1(n_7\to n_7-1)
  \\&
       + (n_2 - 1) I_4(n_7\to n_7-1)
       - 2 (n_2 - 1) I_6(n_7\to n_7-1)
       + (n_2 - 1) I_8(n_7\to n_7-1)
  \\&
       - 2 (n_2 - 1) I_{11}(n_7\to n_7-1)
       + (n_2 - 1) I_{12}(n_7\to n_7-1)
       - 2 (n_2 - 1) I_1(n_9\to n_9-1)
  \\&
       + 2 (n_2 - 1) I_5(n_9\to n_9-1)
       - 4 (n_2 - 1) I_6(n_9\to n_9-1)
       - 2 (n_2 - 1) I_{11}(n_9\to n_9-1)
  \\&
       + 2 (n_2 - 1) I_{12}(n_9\to n_9-1)
  \Bigr]
  \Bigr|_{n_1=n_3=n_4=n_6=1,n_9=0}
  = 0 ,
\end{split}
\end{equation}
which gives
\begin{align*}
  \MoveEqLeft
  \NO(1,n_2,1,1,n_5,1,n_7,n_8,0)
  =
  \frac{1}{(n_2-1)(n_2+n_5+n_8-3+2\ep)}
  \Bigl\{
    \\& - n_7 (2n_2+2n_5+2n_7+n_8-7+4\ep) \NO(1,\GOOD{n_2-1},1,1,\GOOD{n_5-1},1,\BAD{n_7+1},n_8,0)
    \\& - n_8 (2n_5+2n_7+n_8-5+4\ep) \NO(1,\GOOD{n_2-1},1,1,\GOOD{n_5-1},1,n_7,\BAD{n_8+1},0)
    \\& - \bigl[ (n_2^2+4n_2n_5+2n_2n_7+3n_2n_8+4n_5^2+6n_5n_7+4n_5n_8+2n_7^2+3n_7n_8+n_8^2-11n_2-22n_5
    \\& \qquad -15n_7-12n_8+30) + (6n_2+12n_5+8n_7+6n_8-32) \ep + 8 \ep^2\bigr]
    \\& \qquad\quad \times \NO(1,\GOOD{n_2-1},1,1,n_5,1,n_7,n_8,0)
    \\& - (n_2-1)(n_5+n_8-2+2\ep) \NO(1,n_2,1,1,\GOOD{n_5-1},1,n_7,n_8,0)
    \\& + (n_2-1)(n_2+n_5+n_8-3+2\ep) \NO(1,n_2,1,1,n_5,1,\GOOD{n_7-1},n_8,0)
    \\& - (n_2-1)(2n_2+2n_5+2n_7+n_8-7+4\ep) \NO(1,n_2,1,1,n_5,1,n_7,\GOOD{n_8-1},0)
    \\& + n_7 (2n_2+2n_5+2n_7+n_8-7+4\ep) \NO(1,\GOOD{n_2-1},1,\GOOD{0},n_5,1,\BAD{n_7+1},n_8,0)
    \\& + n_8 (2n_2+2n_5+2n_7+n_8-7+4\ep) \NO(1,\GOOD{n_2-1},1,1,n_5,\GOOD{0},n_7,\BAD{n_8+1},0)
    \\& - 2 n_8 (n_2-1) \NO(1,n_2,\GOOD{0},1,n_5,\GOOD{0},n_7,\BAD{n_8+1},0)
    \\& + (n_2-1)(2n_2+3n_5+2n_7+2n_8-9+6\ep) \NO(1,n_2,\GOOD{0},1,n_5,1,n_7,n_8,0)
    \\& + (n_2-1)(3n_2+2n_5+2n_7+2n_8-9+6\ep) \NO(1,n_2,1,1,n_5,\GOOD{0},n_7,n_8,0)
  \Bigr\} .
  \numberthis
\end{align*}
It decreases the total complexity by one and is applicable until $n_1=1$.
Note that, because this rule is used when $n_2 \ge 2$, $n_5 \ge 1$ and
$n_8 \ge 1$, we have $(n_2 + n_5 + n_8 - 3) \ge 1 $ and hence the coefficients
never get any poles.
There exist similar rules for $n_5$, $n_7$ and $n_8$.
Consequently, all integrals in the three-loop non-planar topology are reduced
into one master integral $\NO(1,1,1,1,1,1,1,1,0)$ and integrals in simpler
topologies.


\section{Conclusion}

We have developed a \FORM{} program \Forcer{} for efficient evaluation of
massless propagator-type Feynman integrals up to the four-loop level.
It implements parametric reductions as \Mincer{} does for three-loop integrals.
Due to the complexity of the problem, many parts of the code is generated
in automatic ways.

To check correctness of the program, we recomputed several known results in
the literature, including the four-loop QCD $\beta$-function~%
\cite{vanRitbergen:1997va,Czakon:2004bu} and lower moments of the four-loop
non-singlet splitting functions%
~\cite{Baikov:2006ai,Velizhanin:2011es,Velizhanin:2014fua,Baikov:2015tea}
(see also ref.~\cite{Ruijl:2016pkm}).
Such recomputations also provide practical benchmark tests.
By using the background field method with some tricks~\cite{josLL2016},
the four-loop QCD $\beta$-function was
recomputed within 10 minutes with dropping all gauge parameters and less than
9 hours with fully including gauge parameters (of course the final result is
independent of the gauge parameter), on a decent 24 core machine.
More physics results by \Forcer{} will be reported elsewhere~%
\cite{forcer-physics}.


\acknowledgments

We would like to thank Andreas Vogt for discussions, many useful feedbacks for
the program and collaboration for physics applications.
This work is supported by the ERC Advanced Grant no. 320651, ``HEPGAME''.
The diagrams were drawn with \prog{Axodraw}~\prog{2}~\cite{Collins:2016aya}.

\bibliography{mybibfile}

\end{document}